\documentclass[sigconf]{acmart}

\usepackage{booktabs}
\usepackage{balance}

\setcopyright{rightsretained}

\acmDOI{10.1145/3139937.3139939}

\acmISBN{978-1-4503-5396-0/17/11}

\acmConference{IoT-S\&P'17}{November 3, 2017}{Dallas, TX, USA} 
\acmYear{2017}
\copyrightyear{2017}

\acmPrice{}

\begin{document}
\title{Cleartext Data Transmissions in Consumer IoT Medical Devices}

\author{Daniel Wood}
\affiliation{%
  \institution{Princeton University}
  \city{Princeton} 
  \state{New Jersey} 
}
\email{dewood@princeton.edu}

\author{Noah Apthorpe}
\affiliation{%
  \institution{Princeton University}
  \city{Princeton} 
  \state{New Jersey} 
}
\email{apthorpe@cs.princeton.edu}

\author{Nick Feamster}
\affiliation{%
  \institution{Princeton University}
  \city{Princeton} 
  \state{New Jersey} 
}
\email{feamster@cs.princeton.edu}

\renewcommand{\shortauthors}{D. Wood et al.}

\begin{abstract}
This paper introduces a method to
capture network traffic from medical IoT devices and automatically detect cleartext
information that may reveal sensitive medical conditions and behaviors. 
The research follows a three-step approach involving traffic collection, cleartext detection,
and metadata analysis. 
We analyze four popular consumer medical IoT devices, 
including one smart medical device that leaks sensitive
health information in cleartext. 
We also present a traffic capture and analysis system that seamlessly integrates
with a home network and offers
a user-friendly interface for consumers to monitor and visualize data transmissions
of IoT devices in their homes. 
\end{abstract}




\maketitle

\section{Introduction}

According to the Federal Trade Commission, ``The Internet of Things (``IoT'') refers to the ability of everyday objects to connect to the Internet and to send and receive data'' ~\cite{ftc}. This definition includes a variety of Internet-connected medical devices increasingly deployed in homes and hospital environments. These devices are designed to record patient data and integrate measurements into electronic health records. User data collected by medical IoT devices are especially privacy sensitive, and device manufacturers may be legally obligated to handle such data in accordance with the Health Insurance Portability and Accountability Act of 1996 (HIPAA). 

The Security and Privacy rules of HIPAA require covered entities to maintain appropriate
administrative, technical, and physical safeguards for protecting electronic patient health information~(e-PHI)~\cite{securityHIPAArule, privacyHIPAArule}. HIPAA defines e-PHI as individually identifiable health information, including:
\begin{enumerate}
  \item an individual's past, present or future physical or mental health or condition
  \item the provision of health care to an individual
  \item the past, present, or future payment for the provision of health care to an individual
  \item common identifiers, e.g., name, address, birth date, and  Social Security Number
\end{enumerate}
Entities that collect e-PHI are required to:
\begin{enumerate}
  \item Ensure the confidentiality, integrity, and availability of all e-PHI they create, receive, maintain, or transmit
  \item Identify and protect against reasonably anticipated threats to the security or integrity of the information, and
  \item Protect against reasonably anticipated, impermissible uses or disclosures
\end{enumerate}
\noindent
Many medical IoT devices enable users to track their personal health via their
smartphones and have the potential to leak e-PHI.  Encryption is the most
obvious determinant of confidentiality in medical IoT device communications.
Packets of data sent in the clear can be trivially intercepted by adversaries
and network observers. Even if cleartext data is compressed, it is still
trivial to recover the original content by recovering the compressed message
and attempting decompression using a limited number of widely used compression
algorithms. Transmitting application data in the clear is a severe (and
seemingly obvious) design flaw, and yet it is prevalent among IoT
devices~\cite{tinker}.

Even when medical IoT devices encrypt data transmitted to the cloud, a network
observer could scrutinize metadata to obtain information about a user.
Several recent studies have demonstrated that IoT device traffic analysis can
reveal user behavior  from correlations between device network activity and
user interactions~\cite{apthorpeIoT}.

This paper examines whether in the course of regular behavior, today's
commercially available smart medical devices properly protect all individually
identifiable health information as dictated by HIPAA.  We evaluate a suite of
medical IoT devices, including: (1)~Withings Smart Blood Pressure Monitor,
(2)~Withings Smart Scale, (3)~iHealth Ease Wireless Blood Pressure Monitor, and (4)~1byOne
Digital Smart Wireless Scale, all of which are popular and readily available
on Amazon. We record and analyze network traffic from these devices and
attempt to identify cleartext health information and/or metadata that would
allow an adversary to infer e-PHI from encrypted communications.

We present two main findings. First, we find that multiple devices that were
tested send information related to their users' heath in cleartext. This
result is concerning given our relatively small sample of devices, and
indicates that cleartext transmission of heath information may be a widespread
privacy vulnerability across the market of IoT medical devices. Even more
worrisome, we found cleartext health information in communications from
devices that use SSL/TLS transport layer encryption. This health information
was sent in cookies, URLs, and occasional unencrypted connections---all
indicative of poor development practices by device manufacturers.

Secondly, even the devices that consistently encrypt health information have privacy
vulnerabilities. Network observers and adversaries can see
traffic from these devices as being generated from specific IP addresses. By
examining network traffic from an access point, it is possible to isolate
traffic originating from a fixed set of IP addresses and subsequently mine the
associated metadata for sensitive medical information. Although in some home and
hospital settings, individual IP addresses may be shielded by a NAT, the rise
of IPv6 and devices tethered to cell phones leave many devices directly exposed
and individually identifiable.

Though patients and doctors expect their e-PHI to be properly protected, one
of the major problems with commercially available IoT devices on the market is
the user's lack of visibility in terms of how their data is handled. This
paper presents a simple and intuitive user interface for IoT device users to
monitor and visualize the data that their medical IoT devices transmit to the
Internet. As traffic data from a suite of connected devices is captured on a
local Wi-Fi access point, we mine the traffic for potentially revealing
electronic personal health information that has been transmitted in the clear.
The user interface lists each device connected to the access point, warning
the user of potential cleartext leakages associated with each device (Figure
1). The goal was to make the interface universally intelligible so that
average users with no comprehension of packets, encryption, or networks would
be able to monitor how each device protected their e-PHI.

We recently presented our findings to members of congress, including Senator
Edward Markey~(MA, Committee on Communications and Technology) and Congressman Joe Kennedy~(MA, Subcommittee on Digital Commerce and Consumer Protection), to raise awareness of the
potential vulnerabilites in IoT devices and of the lack of security or privacy
standardization among IoT device manufacturers. As a featured
participant in the Coalition for National Science Funding's 2017 Capital Hill Exhibition,
we addressed the need for increased consumer visibility into the data that their
IoT devices are sending across the Internet through a unified web-based dashboard~\cite{cra}.
\begin{figure}[t]
  \centering
    \fbox{\includegraphics[width=\linewidth]{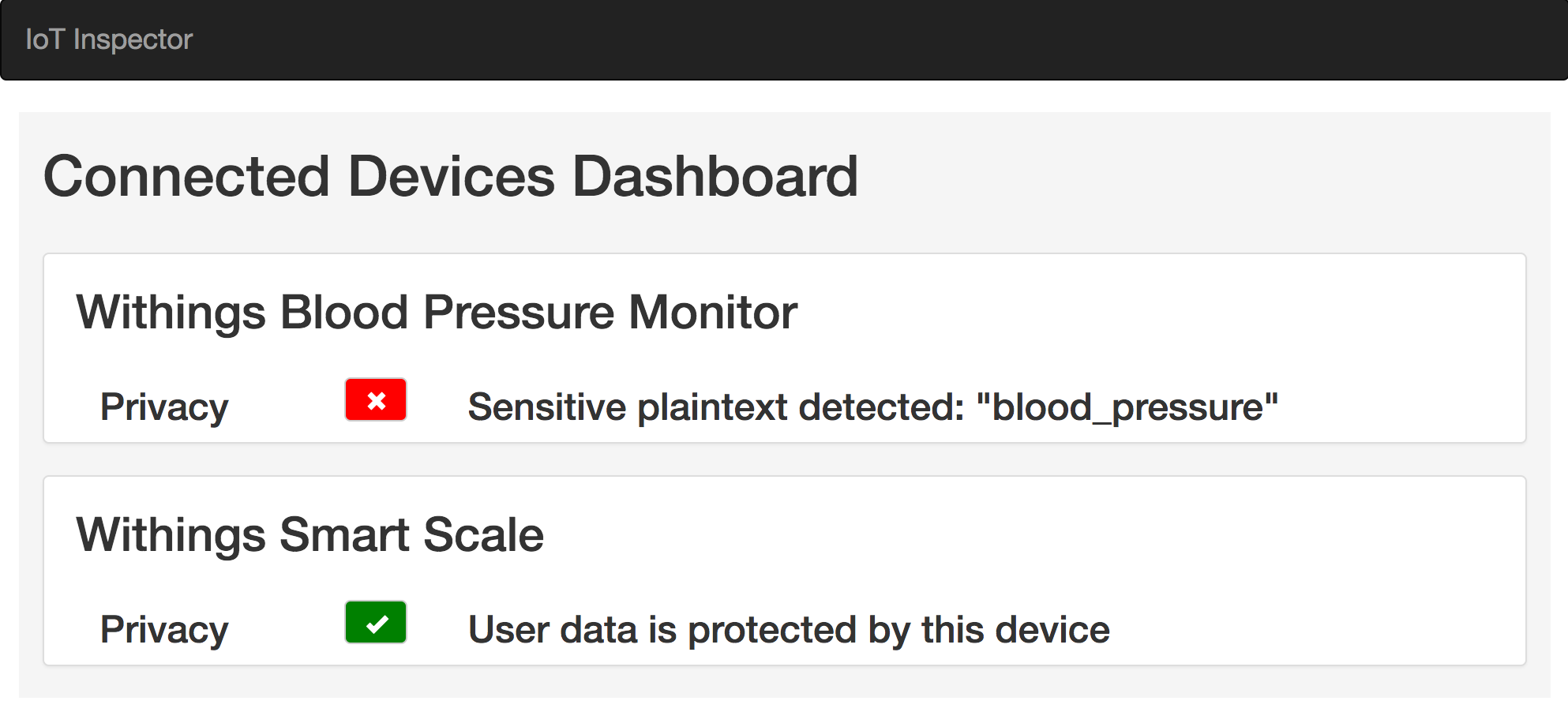}}
  \caption{User interface displays connected devices in the home and privacy status.}
  \label{fig:dashboard}
\end{figure}

\section{Related Work}

Classifying network traffic as
encrypted or cleartext can be challenging. Cha outlines the following
method to determining encrypted versus unencrypted traffic~\cite{chaMachineLearning}:
\begin{enumerate}
  \item Separate each packet's header (non-encrypted) from its payload (potentially encrypted)
  \item Analyze randomness of payload using multiple tests, including Shannon Entropy, Chi-square, and arithmetic mean
  \item Use a training subset from cleartext protocols (HTTP, FTP, Telnet) and encrypted protocols (SSH, TLS) to determine a threshold entropy, above which indicates encrypted traffic
\end{enumerate}
\noindent
We use this approach as a baseline method for classifying traffic. A related
problem is distinguishing encrypted traffic from compressed cleartext traffic,
which is more difficult since both compressed and encrypted traffic exhibit
high orders of entropy. In this paper, we restrict our focus to distinguishing
uncompressed cleartext traffic from encrypted traffic.

Related research into the privacy implications of IoT systems has revealed
significant privacy vulnerabilities that adversaries with passive network
capabilities could exploit. However, most of the literature uses data
collected from generic home devices, not medical IoT devices. For example, Srinivasan et al. present a new privacy
leak in residential wireless ubiquitous computing systems: the Fingerprint and
Timing-based Snooping (FATS) attack~\cite{srinivasan2008fats}.
This attack allows a Wi-Fi eavesdropper to
observe private activities in the home such as cooking, showering, toileting,
and sleeping by snooping on the wireless transmissions of sensors in a home
and leveraging tiered inference algorithms.

 Copos et al. present a scheme to infer when a home is occupied based on parsing
 packet capture files and log characteristics of the network traffic from a smart thermostat~\cite{coposIoT}.
More recently, Apthorpe et al. ~\cite{apthorpeIoT} observed that passive network observers, such as Internet service providers, could analyze IoT network traffic and infer user/device interactions even when device communications are encrypted. 
This attack is especially concerning for personal medical devices. The repetitive nature of medical tests, such as daily blood sugar or blood pressure readings, generates clearly defined patterns of device activity and could reveal common medical conditions from network metadata alone. 

Dimitrov notes that
the proliferation of the medical Internet of Things will revolutionize digital healthcare
by enabling doctors and hospital systems to streamline workflows, increase productivity,
and provide higher data-backed quality of care~\cite{dimitrovIoT}. The research
highlights five key capabilities that leading platforms must enable: (1) Simple
connectivity, (2) Easy device management, (3) Information ingestion, (4) Informative analytics, and (5) reduced risk. In this work, we attempt to improve device management and informative analytics by creating a dashboard that analyzes real-time traffic flows from smart medical devices and informs the user of potential privacy vulnerabilities.

\section{Device Evaluation Methods}

We analyzed medical IoT device network traffic for privacy vulnerabilities using a three phase process:

\begin{enumerate}
  \item Data collection: where we convert a Raspberry Pi into a Wi-Fi access point and collect traffic from a suite of connected IoT devices
  \item Cleartext identification: where we search captured traffic for cleartext application data revealing patient information
  \item Metadata analysis: where we examine second order information such as device activity to infer user behavior
\end{enumerate}

\subsection{Data Collection}

\begin{figure}
  \centering
    \fbox{\includegraphics[width=\linewidth]{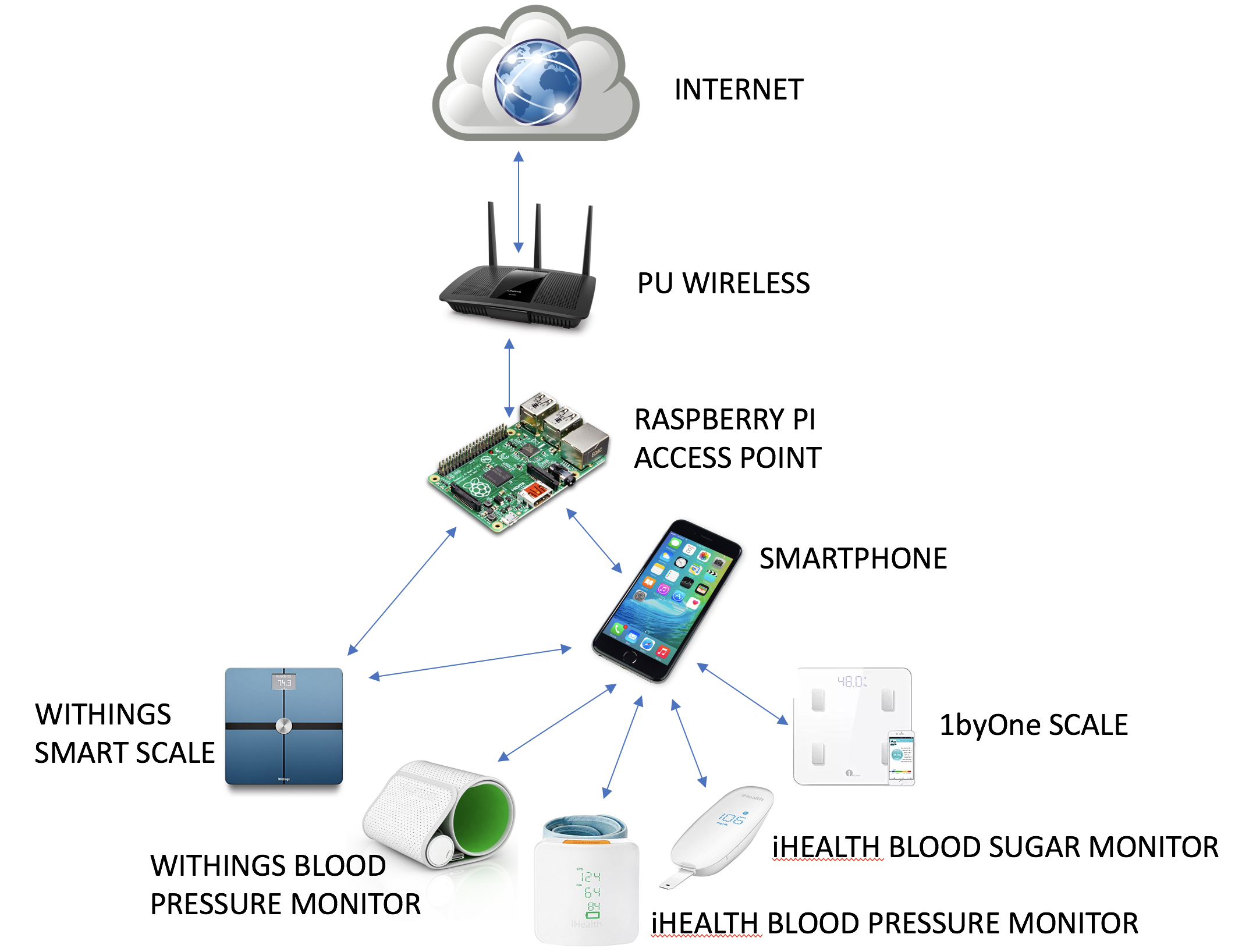}}
  \caption{Data collection environment and connection patterns between devices and infrastructure components.}
  \label{fig:data-collection}
\end{figure}

We created an isolated test environment where we could connect various medical IoT devices to a network and capture live traffic. We configured a Raspberry Pi 3 as a Wi-Fi access point (AP) and programmed it to record traffic to and from connected Wi-Fi stations (Figure~\ref{fig:data-collection}). The open source code can be examined at {\tt github.com/danielwood95/IoTSecurityHub}.  Creating an isolated test environment was necessary because it enabled us to easily separate traffic by device and filter out extraneous traffic on the network.   We chose four medical IoT devices to inspect:

\begin{enumerate}
  \item Withings Wireless Blood Pressure Monitor
  \item Withings Body Composition Wi-Fi Scale
  \item 1byOne Digital Smart Wireless Body Fat Scale
  \item iHealth Ease Wireless Blood Pressure Monitor
\end{enumerate}

We connected the Wi-Fi-enabled devices directly to the Raspberry Pi AP and the Bluetooth-enabled devices to a smartphone connected to the AP.
We purposefully chose two smart scales and two blood pressure monitors so that we could compare the way in which different device manufacturers transmit application data and determine which company had better security or privacy practices, if any. 

We used Wireshark to capture all Ethernet traffic traversing the access point, divided these data using MAC addresses into streams of packets corresponding to each IoT device, and saved the packets in PCAP files for offline analysis. 
 When generating the dataset, we captured as many use cases of each device as possible, including user registration and sign-up, downloading patches and updates, vitals measurements, and health analytics. It is important not to ignore the use cases beyond general vitals measurement.  A malicious adversary could, for example, learn valuable information about the design of a device's embedded system by intercepting a software patch. 

\subsection{Cleartext Identification}

We next analyzed captured packet streams from our medical IoT devices for unencrypted health information. We started by separating the packets by protocol, focusing mainly on HTTP and TCP packets and ignoring packets sent with SSL or TLS, as those are encrypted. Next we separated the payload, which contains application data, from the header of each packet for further analysis. Even though HTTP and TCP are unencrypted, it was still necessary to eliminate payloads containing encrypted application-layer data so we could concentrate our analysis on unencrypted application-layer data. We experimented with three different schemes for this classification: naive ASCII approach, Shannon entropy test, and chi squared test. The latter two are approaches tested by Cha et al. ~\cite{chaMachineLearning}.

\subsubsection{Naive ASCII approach}
If all the characters in the payload are contained within the 128 character ASCII set, we anticipated that a packet would be unencrypted, since encrypted packets would need to contain characters from the extended ASCII set. While the naive ASCII approach does weed out encrypted packets, it does not identify all unencrypted packets, many of which contain characters from the extended ASCII set in addition to the printable characters. 

\subsubsection{Shannon Entropy Test}
The Shannon Entropy Test calculates the entropy of each payload string, which is a quantitative measure of the variability in the frequency of the different possible characters. While random (or in this case, encrypted) strings have very high entropy, unencrypted cleartext and English strings exhibit fairly low entropy. To calculate the Shannon Entropy of a string, let $X$ be a random variable that takes on possible values $x_1$, $x_2$, ..., $x_n$. $p(x_i)$ is the probability that $X = x_i$:
$$H(X) = - \sum_{i = 1}^{n} p(x_i) \log p(x_i)$$
A packet's payload is presumed to be unencrypted if its Shannon entropy value is lower than a threshold parameter. For our analysis, we used a relatively high threshold parameter of $7.5$, so as not to discard some unencrypted payloads with high entropy (at the cost of misidentifying a small number of encrypted payloads with low entropy).

\subsubsection{Chi-Squared Test}
Lastly, the Chi-Squared test compares the observed frequency of each character, $o_i$, with its expected value from a uniform distribution, $e_i$. The value $\chi^2$ is calculated according to the formula:
$$\chi^2 = \sum_{i=1}^{n} \frac{(o_i-e_i)^2}{e_i}$$
The more a set of frequencies deviates from its expected values, the higher the value of $\chi^2$, and if the observed frequencies equal the expected frequencies then $\chi^2$ is 0. Therefore, English cleartext is expected to have a much higher deviation of character frequencies from the expected frequency (a uniformly random distribution). By setting the threshold value $\chi^2 = 1000$, we were able to effectively weed out the unencrypted payloads from those that were encrypted. 

\subsubsection{Method Comparison}

To determine which of the three methods (naive ASCII, Shannon Entropy, or Chi-Squared)
yielded the most accurate classification of unencrypted packets, we ran a comparative analysis with five PCAP files with over 225,000 packets (Figure~\ref{fig:method-comp}). Out of the three methods, the naive ASCII approach was the most selective. It had a 0 percent false positive rate, but missed nearly all of the unencrypted packets, because many of the payloads contained some number of extended ASCII characters. 

In contrast, the Shannon Entropy approach cast a much wider net, tagging a larger share of packets as being unencrypted, but suffered from a high false positive rate. Lowering the threshold entropy value did not significantly increase the accuracy of the approach, as fewer true unencrypted packets were identified as the threshold entropy value decreased. 

We found the most accurate encryption classification approach to be the Chi-Squared
test, due to its low false positive rate and identification of non-random string patterns within the packet payloads we tested. The Chi-Squared test exhibited a false positive rate of approximately 3.5\%, while still identifying nearly all unencrypted payloads detected by the the naive ASCII and Shannon entropy approaches.

\begin{figure}
  \begin{center}
    \begin{tabular}{c|c|c} 
    \textbf{Approach} & \textbf{Precision} & \textbf{\% packets cleartext}\\ [0.5ex] 
    \hline
    Naive ASCII & $1$ & $0.5$ \\ 
    Shannon Entropy &  $0.26$ & $16.2$ \\
    Chi Square & $.97$ & $4.9$ \\
    \end{tabular}
    \caption{Comparison of cleartext detection approaches on 225,000 packet payloads from the devices studied. The precision metric indicates the probability that a particular approach correctly identifies a payload as cleartext. Column~3 indicates what percent of total packets were identified as cleartext. Less selective approaches identify more cleartext packets, but also result in higher false positive rates.}
    \label{fig:method-comp}
  \end{center}
\end{figure}

\subsubsection{Dictionary Analysis}
Once we were able to identify cleartext packets, we identified potentially sensitive personal medical/identifying information by searching each string in the cleartext payload in several dictionaries. We used three dictionaries: a list of the 100 most common medical terms/conditions from Barron's Medical Dictionary~\cite{barrons}, a list of the most popular first male and female names according to the U.S. Census Bureau~\cite{names}, and a list of the most common personal identifying information (i.e. passport number, license, name, address, etc.) according to the National Institute of Standards and Technology~\cite{nist}. 

\subsection{Metadata Analysis}

In cases when devices encrypted application-layer data or utilized secure
protocols such as TLS or SSL, we were still able to infer rough user behavior
during traffic collection due to the fact that the studied devices are typically used to make
periodic measurements and are not always on. For example, blood
sugar may be measured at regular intervals throughout the day (such as after a
meal) and smart scales might be used once every morning. Using Wireshark, we
were able to associate periods of device activity based on time stamps and
origin IP addresses. In some
cases, we were able to determine the user's behavior during these
periods of activity by examining the descriptions of the destination IP
addresses in Wireshark. For example, the Withings Smart Scale always
communicates with {\tt scalews.withings.net} when transmitting data
about a current measurement, making all outgoing traffic easily identifiable.

\section{Device Vulnerability Analysis}

We found a large variability in the methods each device used to send
application data through the network when registering users, sending patches
and updates, measuring vitals, or retrieving health analytics. All of the
devices used encryption and protocols such as TLS or SSH to send sensitive
first order information, such as the user's actual weight or blood sugar
levels. However, there were various degrees of leaking second order
information and metadata, scraped from sources such as HTTP GET requests,
packet header information, and device conversation IP tables. Of the devices
that we captured traffic for, the most secure implementation was the 1byOne
Digital Smart Wireless Body Fat Scale. This device not only used encrypted
protocols to deliver application data, but also masked names of packet
destinations, unlike the Withings devices.

\subsection{Blood Pressure Monitor: \\ Leaks in Cleartext}

The Withings Blood Pressure Monitor, out of the four devices monitored,
exhibited the most vulnerabilities concerning sensitive user information
during data transmission. We were able to capture enough sensitive second-
order data and metadata from network traffic in the course
of typical device use to determine that the user was measuring his or
her blood pressure and how frequently the user was taking these measurements.

First of all, it is easy for a network observer to detect that a Withings IoT
device is in use, because the information sections of all queries and
responses to the Withings servers are titled with the brand of the device in
the URL. This would make it exceptionally easy for a network observer to track
all traffic originating from IP addresses querying an address such as
static.withings.com. Because of the limited capabilities of medical IoT
devices, as opposed to devices such as Amazon Echo, which can reach any
endpoint on the Internet, there is a limited number of endpoints that are
queried from each device, making device identification by a network observer
trivial.

Even more concerning, we observed that one of the signature characteristics of
the Withings Blood Pressure Monitor's traffic pattern was the fact that each
digital reading concluded with a GET request for a stock photo of a person
using the Withings Blood Pressure Monitor (Figure 4). This GET request is
certainly a cause for concern, as any adversary monitoring the traffic would
be able to immediately determine when a user has finished measuring his or her
blood pressure. This GET request was sent completely in the clear, and
furthermore, it is not even displayed on the user interface of the app to the
user of the device. It appears that there is no purpose of sending this image
upon the success of each blood pressure reading, except inadvertently
notifying network observers that the Withings Blood Pressure Monitor is in
use.

\begin{figure}[t]
  \centering
    \fbox{\includegraphics[scale=0.45]{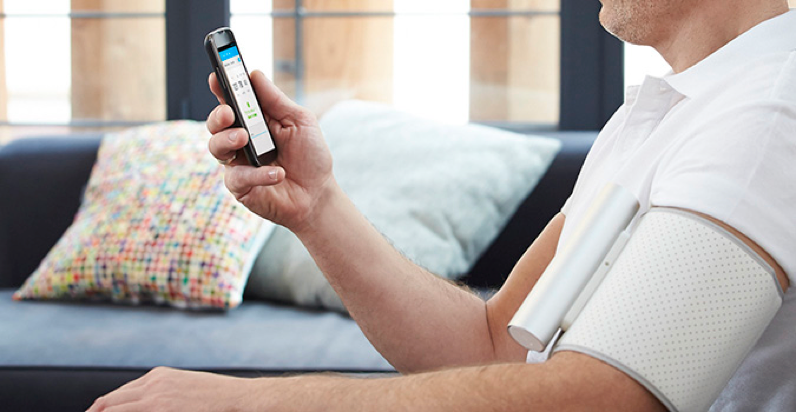}}
     \caption{During data collection phase, the Withings Blood Pressure Monitor sent this image in the clear through a GET request, which reveals the nature of the device traffic.}
     \label{fig:bp-image}
\end{figure}

\begin{figure}[t]
  \centering
    \fbox{\includegraphics[width=\linewidth]{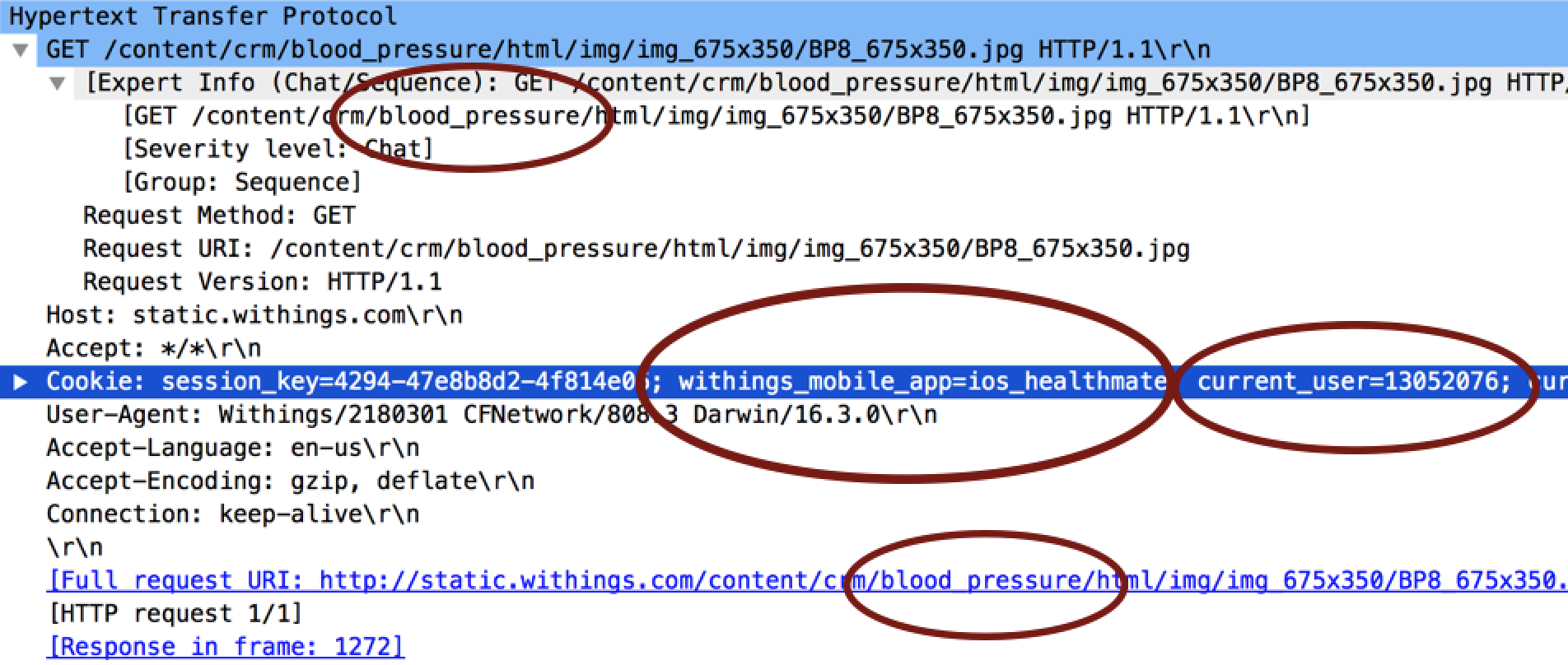}}
  \caption{HTTP packet sent by IoT blood pressure monitor reveals nature of device and user behavior.}
  \label{fig:bp-packet}
\end{figure}

Figure 5 highlights the example of one packet, which alone revealed four sources
of potentially sensitive information about the device. The cleartext string ``blood\textunderscore
pressure'' appears twice, along with the string ``withings\textunderscore mobile\textunderscore app=ios\textunderscore healthmate''. Lastly, the ``current\textunderscore user'' field, while not directly disclosing the name of user, is potentially a unique identifier that associates that user with the subsequent blood pressure data. By monitoring this traffic for a period of time with many users, it would be trivial to match each packet of transmitted application data to the associated user. 

\subsection{Scales and Blood Sugar Monitor: \\ Encryption of User Data}

In contrast to the Withings Blood Pressure Monitor, the Withings and 1byOne
smart scales and iHealth blood pressure monitor did not transmit cleartext
e-PHI. After pairing the devices with a smartphone and connecting them to the
test network to capture the transmitted packets, we found that these devices
actually used TLSv1.2 on port 443 to send encrypted application data.
Additionally, even though the devices only transmitted data when they were
being used to measure weight or blood sugar, the traffic was difficult to
detect without knowing the exact source IP address, since the packets are not
labeled with revealing information about the nature of the device, and the
destination addresses are not readable URLs such as the case of the Withings
Blood Pressure Monitor.

When we ran the deep packet analysis of the traffic, it was not
possible to compile information about the user's behavior in the same way as
with the blood pressure monitor. This suggests that it is relatively easy for
device manufacturers to protect patient information by encrypting
application-layer data and using secure protocols such as TLS or SSL to
transmit application data.

\section{Discussion and Future Work}

The sheer diversity of devices, protocols, and lack of standardization between device manufacturers makes it difficult to detect all vulnerabilities, or to even identify all of the devices that are connected to a network.

This research suggests that medical IoT device manufacturers are not necessarily aligned with policies including the Privacy and Security rules of HIPAA, and they may inadvertently reveal sensitive data and metadata about a user's behavior and medical condition. For example, the HIPAA Privacy rule dictates that manufacturers and medical professionals protect personally identifying information, such as the individual's past, present or future physical or mental health or condition. Though we found no instances of full names or biologically identifiable information being leaked, policy makers and manufacturers should recognize the importance of encrypting all application data and protecting metadata.

This research underscores the lack of awareness among the general public when it comes to the confidentiality and integrity of their personal data. As technology becomes increasingly capable and complex, it will only become more difficult for users of connected devices to comprehend what sort of data can be extracted from their digital footprint, even if the devices they are using encrypt first-order information. Tools like the user interface presented in this paper are in the public interest to increase the visibility of device vulnerabilities, awareness of personal confidentiality weaknesses, and accountability among device manufacturers.

Because the devices examined in this paper are not always on, as in the case
of some other home IoT devices such as an Amazon Echo or Google Nest
Thermostat, future research should examine always-on medical IoT devices, such
as smart glucose pumps. Such devices may have increased demand for security
and privacy, but may additionally make metadata analysis more difficult, since
device traffic is not necessarily correlated with user behavior and activity.

While detecting cleartext application-layer data is an important first step in understanding
the severity of medical IoT security and privacy vulnerabilities, it should be considered "low hanging fruit." Frequently, device manufacturers and software engineers will program IoT devices to transmit payloads that have been compressed in an effort to reduce the number of packets transmitted (and not necessarily as a means of obfuscating cleartext traffic). Thus, an extension of our research would include methods of distinguishing compressed traffic from encrypted traffic. Once this has been done, it would be possible to brute force decompression using a list of widely used decompression algorithms and then apply our deep packet inspection methodology on the resulting cleartext. Distinguishing compressed text (which already has a high entropy value) from encrypted text is not something that the Shannon Entropy or Chi Squared tests are particularly accurate at doing, so a more advanced classification technique, perhaps using a machine learning approach, could be employed. 

\section{Conclusion}

By capturing network traffic from a suite of IoT devices and conducting deep packet inspection, we were able to identify examples of cleartext and metadata leaks that compromise users' privacy. These results reveal multiple known vulnerabilities within IoT devices, but there are heightened implications due to the sensitive nature of the medical metadata being disclosed. If used by healthcare professionals to measure patient vital signs/data over time, these medical IoT devices need to be more carefully examined by regulators and physician networks to increase the awareness of potential privacy violations before they are adopted on a wider scale. 

\section*{Acknowledgments}
Thanks to Rohan Doshi, Gudrun Jonsdottir, and Dillon Reisman. This work was partially supported by a Google Faculty Research Award, NSF Awards CNS-1539902, CNS-1409635, and CNS-1535796, and the Department of Defense through the National Defense Science \& Engineering Graduate Fellowship. 

\bibliographystyle{ACM-Reference-Format}
\balance\bibliography{CleartextIoTData} 

\end{document}